\documentclass[prb]{revtex4}

\usepackage{amsmath, amssymb, esint}

\begin{document}

\title{Gravity at a Quantum Condensate}
%\shorttitle{} %Insert here a short version of the title if it exceeds 70 characters

\author{Victor Atanasov}
\address{Department of Condensed Matter Physics, Sofia University, 5 boul. J. Bourchier, 1164 Sofia, Bulgaria}
\email{vatanaso@phys.uni-sofia.bg}

%\pacs{04.62.+v}{Quantum fields in curved spacetime}
%\pacs{03.65.-w}{Quantum mechanics}
%\pacs{74.50.+r}{Tunneling phenomena (Josephson effects)}

\begin{abstract}

Provided a quantum superconducting condensate is allowed to occupy a curved hyper-plane of space-time,  a geometric potential from the kinetic term arises. An energy conservation relation involving the geometric field at every material point in the superconductor can be demonstrated. The induced three-dimensional scalar curvature is directly related to the wavefunction/order parameter of the quantum condensate thus pointing the way to a possible experimental procedure to artificially induce curvature of space-time via change in the electric/probability current density.


\end{abstract}

%\begin{document}

\maketitle

The General Theory of Relativity (GTR) conveyed an understanding of gravitational phenomena in terms of geometric field\cite{AE}. Its predictions were confirmed in a series of spectacular experiments\cite{P&R, Shapiro, H&T, GPA&B} most notable of which was the detection of gravitational waves by Laser Interferometer Gravitational Wave Observatory (LIGO) and its sister collaboration, VIRGO\cite{LIGO}. 

However, our predicament that gravitation remains an observational phenomenon and not an interaction we can control and artificially create and explore stands. It is the smallness of the coupling constant in GTR $\kappa=8\pi G/c^2 \approx 1.8 \times 10^{-26}$ [m.kg$^{-1}$] that positions us as observers instead of active creators of geometric fields.

Therefore it is of great importance that we actively seek and find whereabouts which allow us to probe the geometric field by artificially creating it. In the present study we circumvent GTR's restriction due to the smallness of the coupling constant by assuming that the geometric field should find a way to express itself in the quantum dynamics of a condensate (superconductor) and by changing the state of the condensate to channel energy in the curvature of the geometric field.  

The manner in which the curvature of space-time, that is the Riemannian space, affects the electronic properties of condensed matters systems on a microscopic (macroscopic in the case of a quantum condensate) scale is largely unknown and due to its experimental accessibility of great interest\cite{C60}. Some previous results in superfluid helium can also be reconsidered as an indication that the quantum condensate can access gravitational degrees of freedom. For example, the thin film of superfluid creeping up the walls of a beaker, that is the Rollin film\cite{Rollin}, only to slowly drop down  lowering its gravitational potential energy can be viewed as an indication that the condensate is sensitive to gravity. Another such instance is the fountain effect, in which the superfluid relaxes thermal energy by converting it to gravitational potential energy\cite{London}.

The framework of constraining quantum particle motion to a curved manifold embedded in a Euclidean space $\mathbb{R}^n$ can be set in two ways: i.) in the {\it intrinsic} quantization approach, the motion is initially constrained to the curved sub-manifold; a  Hamiltonian is constructed from generalized coordinates and momenta intrinsic to the sub-manifold and then quantised canonically; in this approach, the embedding space is inaccessible and the quantum system depends only on the geometry intrinsic to the sub-manifold\cite{Leaf, deWitt, Birrell&Davies, Wald}; 
ii.) in the {\it confining} potential approach, a free in the embedding space quantum particle is subjected by a normal to the sub-manifold force that gradually confines the dynamics onto it; the Hamiltonian depends on the intrinsic geometry and on the way this sub-manifold
is immersed in the embedding space\cite{Jensen&Koppe, daCost*81, ogawa, Goldstone&Jaffe, Exner*01, Schuster&Jaffe}.

The present paper explores not only the possibility of using superconducting condensates to probe the curvature of space-time, but more importantly to create curvature of space-time by changing the state of the condensate. In order to present our argument we will make the clarification that the quantum motion will not be constrained onto a curved manifold but rather allow the space-time embedding of the condensate to be curved and look for relationship between the curvature and the condensate's wavefunction. Therefore, we will assume the condensate be sensitive to the actual curvature of space-time (or the spatial part of it), that is, the condensate can feel and couple to the gravitational field in a way allowing channelling of energy from the quantum system to the geometric field.
The non-relativistic quantum mechanics of a condensate in a three dimensional hyper-plane of the four dimensional curved space-time can be treated in a well established manner.  Here we suppose that the space-time which the condensate abides is curved. Suppose the four dimensional space-time $M$ is topologically the product $M\cong \mathbb{R}\times \Sigma,$ where $\Sigma$ represents a space-like three dimensional hyper-plane. We can then foliate $M$ by a one parameter family of embeddings given by the map $\tau_t:$ $\Sigma \to M$ such that $\Sigma_t=\tau_t(\Sigma) \subset M,$ that is $\Sigma_t$ is the image of the map $\tau$ in $M$ for a fixed ''time'' $t.$ We assume that the leaves $\Sigma_t$ are space-like with respect to the metric in $M.$ As a result, there exists in $M$ a time-like field normal to the leaves $\Sigma_t.$ In effect, the space-time is splittable into 3+1 dimensions and the induced Riemannian metric $g_{ij}$ onto the three dimensional $\Sigma_t$ can be used to write the Laplace-Beltrami operator, which is the kinetic energy term in the Schr\"odinger equation for the subjected to the geometric field quantum condensate.

The emergence of the geometric field from the kinetic term can be made clearer in the vicinity of the origin where the following Taylor expansion of the induced metric in normal coordinates applies: $g_{ij}=\delta_{ij}-\frac13 R_{ikjl}x^{k} x^{l} + O(|x|^3)$ \cite{Viaclovsky}. 
Using a standard re-normalisation of the wave-function $\Psi=\psi/|g|^{1/4}$ and keeping the lowest order terms (the only relevant for the quantum dynamics) in the Taylor expansion we arrive at the Schr\"odinger equation
\begin{eqnarray}\label{Schrodinger&R}
\nonumber i\hbar \partial_t \psi &=&\frac{1}{2m}\left(\frac{\hbar}{i} \nabla - q \vec{A}  \right).\left(\frac{\hbar}{i} \nabla - q \vec{A}  \right) \psi \\
&&+\; qU\psi+ V_{Geom}  \psi. 
\end{eqnarray}
Here $V_{Geom}=- \frac{\hbar^2}{2m} \alpha R, 
$ where $R$ is the three-dimensional Ricci scalar curvature and $\alpha=1/12$ is a numeric coefficient. Note, $\Delta$ is the Laplacian on flat space. The electric field is defined with the potential $U$ and the magnetic field, defined through the vector potential $\vec{A}.$ 
For a more detailed derivation and discussion, refer to \cite{Victor}.

Next, let us assume we have a Cooper pair condensate inside a superconductor at hand. The Schr\"odinger equation for the quantum condensate will be (\ref{Schrodinger&R}) with $q=2e,$ that is twice the charge of the electron. This equation will describe the state of the entire condensate. Therefore, we may write  $\psi=|\psi(\vec{r})|e^{i\theta(\vec{r})},$ where $|\psi|^2$ is the the charge density of the condensate and $\theta(\vec{r})$ its phase. Upon substitution of this trigonometric form of the wave-function into (\ref{Schrodinger&R}) we can separate the real and imaginary part of the equation to arrive at slightly modified standard result:
\begin{eqnarray}\label{eq:rho_t}
\frac{\partial |\psi|^2}{\partial t}&=& - \nabla. \vec{J}, \\\label{eq:J}
\vec{J}&=&\vec{v}|\psi|^2=\frac{1}{m}\left( \hbar \nabla \theta - q \vec{A} \right) |\psi|^2,\\\label{eq:theta_t}
\nonumber \hbar \frac{\partial \theta}{\partial t}&=&qU-\frac{1}{2m} \left( \hbar \nabla \theta - q \vec{A} \right)^2 \\
&&+ \frac{\hbar^2}{2m} \left( \frac{\Delta |\psi|}{|\psi|} + \alpha R \right).
\end{eqnarray}
Here $\vec{J}$ is the current density, which in the case of a superconducting condensate stands also for the probability density current. The generalised momentum is contained in the expression for$\vec{J}:$ $\vec{p}= \hbar \nabla \theta - q \vec{A},$ therefore the current density is just the velocity of the superconducting current times the charge density.

These two equations are the equations of motion of the superconducting Cooper pair fluid in the presence of an induced from the embedding space-time curvature $R.$

Upon formal integration with respect to the temporal coordinate, we can rewrite the continuity equation (\ref{eq:rho_t}) which expresses  the current/probability density conservation in the following way:
\begin{eqnarray}\label{eq:continuity}
|\psi|= \sqrt{|\psi_0|^2- \int^t {\rm div} \vec{J} d\tau }, 
\end{eqnarray}
where $|\psi_0|$ is the wavefunction's amplitude at $t=0.$ Note the choice of the integration constant is determined by the initial condition. For the sake of the goal we have in mind, that is the demonstration of a necessary condition to induce an artificial geometric field, we assume that in the initial moment the condensate is at rest and $|\psi(t=0)|= |\psi_0|.$

Now, let us go back to (\ref{eq:theta_t}) which is the hydrodynamic form of the Schr\"odinger equation coding the condensate dynamics. This equation describes the standard superconducting phase evolution
$\hbar \frac{\partial \theta}{\partial t}=2eU,$ provided $\hbar \nabla \theta = q \vec{A}, $ that is the current density in the bulk vanishes and a new requirement which stems from the properties of the geometric potential applies:
\begin{equation}\label{eq:R_psi}
R=-\frac{1}{\alpha} \frac{\Delta |\psi|}{|\psi|}.
\end{equation}
Note once again that $R$ is the three-dimensional induced scalar curvature. It is related to the four-dimensional Ricci curvature of space-time with the relation $R^{4D}=\frac{4}{3} R^{3D}$ (see \cite{Ficken}). This relation is a product of our initial assumption that the quantum condensate can abide a curved space-time.

In the paper we claim that it is the spatial variation of the wavefunction that can affect the scalar curvature.  Althought, the procedure behind (\ref{Schrodinger&R}) seems like a confining  procedure that might lead to curvature causing spatial variation of the wavefunction amplitude, it is not. Confining procedures can destroy the hydrodynamic interpretation of the Schr\"odinger equation since it is strictly 3+1 dimensional. Equation (\ref{Schrodinger&R}) is a form of a Taylor series around the flat metric, that is to say, the effect of curvature to first order. It then enters the hydrodynamic interpretation of the Schr\"odinger equation and acts similarly to the Bohm's quantum potential ; Bohmian interpretation is more applicable than the confining one. Since the wavefunction has a probabilistic interpretation, literally it is an information field, the Bohmian interpretation actually points to gravitational field having the same nature.

The cornerstone relation (\ref{eq:R_psi}) is the main result of the present study. It emerges from the requirement that the condensate phase evolution remains the one which is experimentally well verified in the Josephson junction tunnelling device\cite{JJ}. However, it might be rather unclear why the state of the condensate would be able to generate curvature. The proposition made here, requires experimental verification but before such an attempt is made we turn our attention to (\ref{eq:J}). Although it is likely that the curvature will modify the dynamics of the condensate, this is not the case. The geometry of the embedding space-time does not affect the canonical momentum (\ref{eq:J}), that is the external to the condensate gravitational field is not doing work and not changing condensate's energy. Therefore, we hypothesise that the energy change as the condensate changes state should channel somewhere and one possibility being the geometric field as (\ref{eq:R_psi}) governs. We believe this is a very specific mechanism that imposes the curvature to adapt to the state of the condensate, a mechanism that we made explicit and justify equation (\ref{eq:R_psi}).

Yet, another argument to the validity of (\ref{eq:R_psi}) is the empirically semi-classical behaviour of superconductors, besides the common macroscopic quantum state all Cooper pairs share, the dynamics of the charged fluid is strictly classical, that is the presence of the Bohmian-like quantum force $\frac{\hbar^2}{2m} \left( \frac{\Delta |\psi|}{|\psi|} + \alpha R \right)$ is not felt. In the case of a superconductor in flat space, the standard requirement to account for the lack of empirical presence of the quantum force is the homogeneity of the wave function that is
\begin{eqnarray}
\Delta |\psi| = 0 \quad \Rightarrow \quad \frac{\hbar^2}{2m} \left( \frac{\Delta |\psi|}{|\psi|}\right)=0.
\end{eqnarray}
Therefore, we explore the case in which the Bohmian-like quantum force vanishes in the curved-space time case as well.

Note, (\ref{eq:R_psi}) does not mean that there is no condensate in flat space where $R=0,$ but rather $\Delta |\psi|=0$ which is easily fulfilled in all practical experimental cases where condensate's density is assumed constant\cite{feynman}. Not only is $\Delta |\psi|=0$ easily fulfilled but rather required in order to produce the standard superfluid behaviour of the condensate. Now, in curved space-time the standard superfluid behaviour can be restored only if (\ref{eq:R_psi}) applies. Therefore, it is of paramount importance for the verification of the proposed effect to force the condensate into a state of 
$$\Delta |\psi| \neq 0.$$

Now, we apply the continuity equation (\ref{eq:continuity}) to (\ref{eq:R_psi}) to obtain:
\begin{eqnarray}\label{eq:R_J}
\nonumber R=&&\frac{1}{2\alpha} \;\frac{ \Delta \int^t {\rm div} \vec{J} d\tau }{ |\psi_0|^2- \int^t {\rm div} \vec{J} d\tau} \\
&& - \frac{1}{4\alpha}  \frac{ \left({\rm grad} \int^t {\rm div} \vec{J} d\tau \right)^2 }{\left( |\psi_0|^2- \int^t {\rm div} \vec{J} d\tau \right)^2 }.
\end{eqnarray}
The expression is rather cumbersome but reveals the relation of the scalar curvature $R$ to an experimentally controllable parameter, that is the current density $\vec{J}$ in the case of a superconductor. The curvature is related to the spacial change in the supercurrent flux, that is ${\rm div} \vec{J},$ in a straight-forward manner. {\it No coupling constants are involved.} However, the dynamics of the geometric perturbation cannot be explored in the present framework.

The above expression (\ref{eq:R_J}) can reveal different controllable mechanisms to artificially induce curvature of space-time, but we will focus on a single mechanism which holds the greatest of promise. We would require the vanishing of the  denominator, which would lead to a divergence in the induced three-dimensional scalar curvature
\begin{equation}\label{eq:R_infty}
\lim_{ |\psi_0|^2 \to \int^t {\rm div} \vec{J} d\tau  } R = \infty.
\end{equation}
In effect, we would like to focus on the experimental possibility to set
\begin{equation}\label{}
\int^t {\rm div} \vec{J} \; d\tau = |\psi_0|^2.
\end{equation}
Here $|\psi_0|^2$ is the initial quantum condensate density (charge density in the case of a superconductor). Suppose we volume integrate over a three dimensional domain $V$ with the boundary $\partial V =S$ and apply Gauss's theorem to arrive at a clearer meaning
\begin{equation}\label{}
\int^t \left(  \iiint_V  {\rm div} \vec{J} dV \right) d\tau= \int^t \left(  \oiint_S  \vec{J}. d\vec{S} \right) d\tau = |\psi_0|^2 V.
\end{equation}
One can read the above mathematical statement as 
\begin{eqnarray*}
{\it the \; total \; supercurrent \; outflux } = \\
{\it total \; number \; of \; quantum \; condensate \; pairs}, 
\end{eqnarray*}
which condition is easily met during the process of condensate destruction. 

The divergence of the scalar curvature during the process of condensate destruction seems to violate energy conservation. However, this is not the case as the following energy conservation applies \cite{Victor} 
\begin{eqnarray}\label{eq:energy_conservation}
W_g(\vec{r}) - \mathcal{E}(\vec{r}) +  W_{int}(\vec{r}) ={\rm const}.
\end{eqnarray}
Here $\mathcal{E}(\vec{r})=2eU$ is the electrical energy of the Cooper pairs, $W_{int}(\vec{r})=q \vec{v}.\vec{A}$ is the energy of the interaction between the charged  quantum condensate and the magnetic field and $W_g(\vec{r})=\alpha \frac{\hbar^2}{2m}  R (\vec{r})$ is the energy of the geometric field. The energy conservation relation can be derived by taking a gradient of (\ref{eq:theta_t}), substituting the relevant quantities from (\ref{eq:rho_t}) and then integrating along a trajectory. This energy conservation relation applies in every material point and serves as a guidance to the amount of curvature one might expect to create during an electric discharge through a Josephson junction and the accompanying it condensate destruction. In the particular case of $W_{int}(\vec{r})={\rm const},$ the electrical energy of the discharge can be converted into curvature
$2e \delta U=\alpha \frac{\hbar^2}{2m}  \delta R.$ In effect, the Josephson junction can act as a voltage-to-curvature converter.

%\section{Conclusion}

To summarise, the origin of the  geometric potential is in the kinetic term of a constrained to a curved three-dimensional hyper-plane of space-time quantum mechanical condensate (superconductor). This potential manifests in the hydrodynamic interpretation of the Schr\"odinger equation in a way similar to the Bohm's "quantum potential." When external electromagnetic field is included in the dynamics and suitable simplifications applied, one is able to derive an energy conservation relation at every material point in the superconductor. This conservation relation includes a geometric field part associated with the curvature of the hyper-plane. 

When one requires the standard and experimentally relevant quantum dynamics for the phase of the condensate, a relation between the induced three-dimensional scalar curvature and the condensate wavefunction emerges. This relation points to a possible mechanism of artificial geometric field creation, that is creation of curvature of space-time. The most promising approach is the destruction of the condensate (for example, in an electric discharge at a Josephson junction, a superconducting tunnelling device in which a voltage drop can be developed). The present study represents a guidance to a future experimental effort to verify the effect.

%\begin{thebibliography}{0}

%\bibitem{b.a}
 % \Name{Author F., Author S. \and Author T.}
  %\REVIEW{Some Rev. A}{69}{1969}{9691}.

%\bibitem{b.b}
 % \Name{Author F. \and Author S.}
  %\Book{Some Book of Interest}
  %\Editor{A. Editor}
 % \Vol{9}
 % \Publ{Publishing house, City}
  %\Year{1939}
  %\Page{666}.

%\bibitem{b.c}
 % \Editor{Editor A.}
 % \Book{Some Book of Interest}
 % \Vol{9}
  %\Publ{Publishing house, City}
  %\Year{1939}
  %\Section{A}.

%\end{thebibliography}

\end{document}